\documentstyle[12pt,epsf]{article}
\newlength{\abstractwidth}
\renewcommand{\thefootnote}{\fnsymbol{footnote}}
\renewcommand{\thanks}[1]{\footnote{#1}} % Use this for footnotes
\newcommand{\starttext}{
\setcounter{footnote}{0}
\renewcommand{\thefootnote}{\arabic{footnote}}}
\renewcommand{\theequation}{\thesection.\arabic{equation}}
\newcommand{\be}{\begin{equation}}
\newcommand{\bea}{\begin{eqnarray}}
\newcommand{\eea}{\end{eqnarray}}
\newcommand{\beq}{\begin{equation}}
\newcommand{\ee}{\end{equation}}
\newcommand{\eeq}{\end{equation}}

\renewcommand{\a}{\alpha}

\def\ba{\begin{eqnarray}}
\def\ea{\end{eqnarray}}

\def\12{{1 \over 2}}
\def\32{{3 \over 2}}
\def\72{{7 \over 2}}
\def\92{{9 \over 2}}

\def\a{\alpha'}
\def\nc{non-commutative}

\begin{document}
\renewcommand{\theequation}{\thesection.\arabic{equation}}
\begin{titlepage}
\bigskip
\hskip 3.7in\vbox{\baselineskip12pt
\hbox{}
\hbox{SU-ITP 00-09}
\hbox{hep-th/0003075}}

\bigskip\bigskip\bigskip\bigskip

\centerline{\Large \bf 
Invasion of the Giant Gravitons from Anti de Sitter Space}

\bigskip\bigskip
\bigskip\bigskip
\centerline{ John McGreevy, Leonard Susskind and Nicolaos Toumbas
\footnote[1]{\tt mcgreevy@stanford.edu, susskind@sewerrat.stanford.edu} }
\bigskip
\bigskip
\centerline{\it Department of Physics}
\centerline{\it Stanford University}
\centerline{\it Stanford CA 94305-4060}
\bigskip\bigskip
\begin{abstract}

\medskip
\noindent It has been known for some time that the AdS/CFT
correspondence predicts a limit on the number of single particle
states propagating on the compact spherical component of the $AdS \times S$
geometry. The limit is called the stringy exclusion principle. The
physical origin of this effect has been obscure but it is usually
thought of as a feature of very small distance physics.
In this paper we will show that the stringy exclusion principle is
due to a surprising large distance phenomenon. The massless single
particle states become progressively less and less point-like as
their angular momentum increases. In fact they blow up into
spherical branes of increasing size. The exclusion principle is
simply understood as the condition
that the particle should not be
bigger than the sphere that contains it.
\end{abstract}
\end{titlepage}
\starttext
\baselineskip=18pt
\setcounter{footnote}{0}

%%%%%%%%%%%%%%%%%%%%%%%%%%%%%%%%%%%%%%%%%%%%%%%%%%%%%%%%%%%%%%%%%%%%%%
%%%%%%
%%%%%%%%%%%%%%%%
%%%%%%%%%%%%%%%%%%%%%%%%%%%%%%%%%%%%%%%%%%%%%%%%%%%%%%%%%%%%%%%%%%%%%%
%%%%%%
%%%%%%%%%%%%%%%%
\setcounter{equation}{0}
\section{Introduction}
In conventional (20th
century) physics, high energy or high momentum came to be
associated with small distances. The physics of the 21st century
is likely to be dominated by a very different perspective.
According to the
Infrared/Ultraviolet connection \cite{1} which underlies much of
our new understanding of
string theory and its connection to gravity, physics of increasing energy
or momentum
is governed by
increasingly large distances. Examples include the growth of
particle size with momentum \cite{2} \cite{3}, the origin of the
long-range gravitational
force in Matrix Theory
from high energy quantum corrections \cite{4} and the IR/UV connection in AdS
spaces. Another important
manifestation is the spacetime uncertainty principle of string
theory \cite{5} \cite{6} \cite{7}
\be
\Delta x \Delta t \sim \a.
\ee
Similar uncertainty principles occur in non-commutative geometry
where the coordinates of space do not commute. An important
consequence of the non-commutativity is the fact that the particles
described by non-commutative field theories have a spatial
extension which is proportional to their momentum \cite{8}\cite{9}. This in
turn
leads to unfamiliar violations of the conventional decoupling of
IR and UV degrees of freedom in these theories \cite{10}\cite{11}\cite{12}.
In this paper we will describe another example of IR/UV
non-decoupling that occurs in AdS/CFT theories. The relevant
space-times have the form $AdS_n \times S^m$. We are interested in
the motion of the graviton and other massless bulk particles on
the $S^m$. The motion is characterized by an angular momentum $L$
or more exactly a representation of the rotation group $O(m+1)$.
In 20th century physics such particles are regarded as point or
almost point particles regardless of $L$. In fact we will see that
as $L$ increases the particles blow up in size very much like the
quanta of non--commutative field theories. When the size reaches the radius of
the $S^m$,
the growth can no longer continue and
the tower of Kaluza--Klein states terminates. This is the
origin of the stringy exclusion principle \cite{13}\cite{14}\cite{15}.

In section 2 we will review the theory of electric dipoles moving in a
magnetic field. This system is the basic object of \nc \ field
theory. When the theory is defined on a 2-sphere there is a bound
on the angular momentum when the ends of the dipole separate to
the antipodes of the sphere. In sections 3, 4 and 5 we consider the
cases of $AdS_7 \times S^4$, $AdS_4 \times S^7$ and $AdS_5 \times
S^5$. In each case we find that the spectrum of angular momentum is bounded
and that the bound agrees with expectations from the stringy
exclusion principle.

%%%%%%%%%%%%%%%%%%%%%%%%%%%%%%%%%%%%%%%%%%%%%%%%%%%%%%%%%%%%%%%
%%%%%%%%%%%%%%%%%%%%%%%%%%%%%%%%%%%%%%%%%%%%%%%%%%%%%%%%%%%%%%%
%%%%%%%%%%%%%%%%%%%%%%%%%%%%%%%%%%%%%%%%%%%%%%%%%%%%%%%%%%%%%%%

\setcounter{equation}{0}
\section{Dipoles in Magnetic Fields }
In this section we briefly review the dipole analogy for \nc \ field
theory \cite{8} \cite{9}.
We begin with a pair of unit charges of opposite sign
moving on a plane with a constant magnetic field $B$. The
Lagrangian is
\be
{\cal L} ={m \over 2} \left(\dot x_1^2 +\dot x_2^2 \right)
+{B\over 2} \epsilon_{ij} \left(\dot x_1^i x_1^j - \dot x_2^i x_2^j \right)
-{K \over2}(x_1 -x_2)^2.
\ee
Let us suppose that the mass is so small so that the first term in
Eq.(2.1) can be ignored. Let us also introduce center of mass and
relative coordinates
\bea
X&=&(x_1 +x_2)/2 \cr
\Delta &=& (x_1 -x_2)/2.
\eea
The Lagrangian becomes
\be
{\cal L}=B \epsilon_{ij}\dot X^i \Delta^j -2K\Delta^2.
\ee
From Eq.(2.3) we first of all see that $X$ and $\Delta$ are
non-commuting variables satisfying
\be
[X^i ,\Delta^j]=i{\epsilon^{ij}\over B}.
\ee
Furthermore the center of mass momentum conjugate to $X$ is
\be
P_i=B\epsilon_{ij}\Delta^j.
\ee
Thus when the dipole is moving with momentum $P$ in some direction
it is stretched to a size
\be
|\Delta| =|P|/B.
\ee
in the perpendicular direction.
This is the basis for the peculiar non-local effects in non--commutative field
theory.

Now suppose the dipole is moving on the surface of a sphere of
radius $R$. Assume also that the sphere has a magnetic flux $N$.
In other words there is a magnetic monopole of strength
\be
2 \pi N= \Omega_2 B R^2.
\ee
at the center of the sphere. We can get a rough idea of what happens
by just saying that when the momentum of the dipole is about $2BR$
the dipole will be as big as the sphere. At this point the angular
momentum is the maximum value
\be
L=PR \sim BR^2.
\ee
This is of order the total magnetic flux $N$.

We will now do a more precise analysis and see that the maximum angular
momentum is exactly $N$. Parameterize the sphere by two angles $\theta,\phi$.
The angle $\phi$ measures angular distance from the equator. It is
$\pm \pi/2 $ at the poles. The azimuthal angle $\theta$ goes from
$0$ to $2\pi$. We work in a gauge in which the $\theta$ component
of the vector potential is non-zero. It is given by
\be
A_{\theta} = N{1-\sin{\phi}\over 2R \cos{\phi}}
\ee
For a unit charged point particle moving on the sphere the term coupling the
velocity to the vector potential is
\be
{\cal L}_A =A_{\theta} R \cos{\phi}\dot{\theta}=N R{1-\sin{\phi}\over 2R }
\dot{\theta}.
\ee

Now consider a dipole with its center of mass moving on the
equator. The positive charge is at position $(\theta,\phi)$ and the
negative charge is at $(\theta,-\phi)$. For the motion we
consider $\phi $ is time independent. Eq.(2.10) becomes
\be
{\cal L}_A =N({1-\sin{\phi} \over 2})
\dot{\theta} - N({1 + \sin{\phi} \over 2})
\dot{\theta}
\ee
or
\be
{\cal L}_A ={-N\sin{\phi}}
\dot{\theta}.
\ee

Again we want to consider a slow-moving dipole whose mass is 
so small that its kinetic term may be ignored 
compared to the coupling to the 
magnetic field, i.e. $mR << N$.
Let us also add a spring coupling 
\be
{\cal L}_S= - {k \over 2} R^2 \sin^2 \phi;
\ee
for simplicity, we have used the chordal distance in 
this potential.
The total Lagrangian is
\be
{\cal L}=  - {k \over 2} R^2 \sin^2 \phi
- N \sin{\phi}\dot{\theta}
\ee
and the angular momentum is
\be
L = - N\sin{\phi}.
\ee
The angular momentum will reach its maximum when $\phi = \pi/2$ at
which point
\be
|L_{max}|= N.
\ee
The fact that the angular momentum of a single field quantum in \nc\ field
theory is
bounded by $N$ is well known in the context of non--commutative field theory
on a
sphere \cite{20}. Here we see that it is a large distance 
effect.

%%%%%%%%%%%%%%%%%%%%%%%%%%%%%%%%%%%%%%%%%%%%%%%%%%%%%%%%%%%%%%%
%%%%%%%%%%%%%%%%%%%%%%%%%%%%%%%%%%%%%%%%%%%%%%%%%%%%%%%%%%%%%%%
%%%%%%%%%%%%%%%%%%%%%%%%%%%%%%%%%%%%%%%%%%%%%%%%%%%%%%%%%%%%%%%

\setcounter{equation}{0}
\section{$AdS \times S$}

We now study BPS particles moving on the sphere of maximally supersymmetric
$AdS$ vacua of string and M theory.  For simplicity, we present all
details of the argument in the case of the M5-brane geometry.  Results
for the other cases are given in the latter subsections.  Note that in all
of these cases, the energy of our objects is well below the
energy of a stable AdS black hole.

\subsection{$AdS_7 \times S^4$}
We are interested in the motion of a BPS particle
on the 4-sphere of $AdS_7 \times
S^4$. We will assume that the radius of curvature $R$ is much larger than the
11 dimensional Planck length $l_p$. The analogy with the previous example
is very close. The
role of the magnetic field is played by the 4-form field strength
on the sphere. We call the flux density $B$. Quantization of flux requires
\be
\Omega_4 BR^4 = 2 \pi N.
\ee
From the supergravity equations of motion it can be seen \cite{17}
that $R$ is given by
\be
R=l_p {(\pi N)}^{1\over3}.
\ee
The assumption of large $R$ means $N>>1$.

We want to know what happens to a graviton or any other massless 11
dimensional particle when it moves on the 4-sphere in the presence
of the 4-form field strength. As long as the angular momentum is
small ($L<<N$) the graviton is expected to be much smaller
than $R$, and we can make the approximation that space is locally
flat. Furthermore from Eq.(3.1) we see that the field strength $B$
is small
\be
B \sim N^{-{1\over 3}}l_p^{-4}.
\ee

Since the space is almost flat we can locally introduce flat space
coordinates $x^0,....,x^{10}$. Let us take the $B$ field to lie in
the $(7,8,9,10)$ directions. The graviton is moving along the
$x^{10}$ direction. Its momentum is $P_{10} = L/R$. This setup is
very close to one that was studied by Myers using Matrix Theory
\cite{16}.

In matrix theory the 11 dimensional graviton is viewed as a
threshold bound state of $n=P_{10}R_{10}$ D0-branes. Myers shows that
in a background 4-form fieldstrength the D0-brane configuration is described
as a spherical membrane with a radius $r$ that grows with $P_{10}$
according to
\be
r  \sim BP_{10}l_p^6.
\ee
Let us assume that this formula is approximately valid until $r$
becomes of order $R$. In that case when the graviton size becomes $\sim R$
it will have momentum
\be
P_{10}(max) \sim R/{B l_p^6}
\ee
and angular momentum
\be
L_{max} \sim RP_{10}(max) \sim R^2/{B l_p^6} \sim N.
\ee
Thus as in the previous section we find that the maximum single
particle angular momentum is $N$.

We will now give a more precise calculation which parallels that of section 1.
We are interested in the dynamics of a relativistic spherical membrane
moving in $S^4$.
The membrane has zero net charge but it couples to the background
field strength.
It behaves like the dipole of section 1.

Let us parametrize $S^4$ using cartesian coordinates $X_1,.....,X_5$ so that
\bea
X_1& =& R \cos \theta_1 \cr
X_2& =& R \sin \theta_1 \cos \theta_2 \cr
X_3& =& R \sin \theta_1 \sin \theta_2 \cos \theta_3 \cr
X_4& =& R \sin \theta_1 \sin \theta_2 \sin \theta_3 \cos \theta_4 \cr
X_5& =& R \sin \theta_1 \sin \theta_2 \sin \theta_3 \sin \theta_4.
\eea
The angles $\theta_1,....,\theta_3$ go from $0$ to $\pi$. The angle
$\theta_4$ is the azimuthal angle and goes from $0$ to $2 \pi$.
Then
\be
X_1^2 + X_2^2 + X_3^2 + X_4^2 + X_5^2 = R^2.
\ee

Next we embed a spherical membrane in $S^4$. We choose to parametrize
the surface of the membrane by $\theta_3, \theta_4$. The brane is
allowed to move in the $X_1, X_2$ plane. Its size depends on its
location in the $X_1, X_2$ plane according to
\be
r = R\sin\theta_1\sin\theta_2.
\ee
We see that when the size is at its maximum value, $r = R$, the
membrane is at the origin $X_1 = X_2 = 0$ like the two charges at the
ends of the dipole in section 1. Since
\be
X_1^2 + X_2^2 = R^2 - r^2,
\ee
the membrane can move around a circle in the plane and have constant size.
We also set
\bea
X_1 &=& \sqrt{R^2 - r^2}\cos\phi \cr
X_2 &=& \sqrt{R^2 - r^2}\sin\phi.
\eea
In terms of the coordinates $r, \phi, \theta_3, \theta_4$, the
metric on the 4--sphere becomes
\be
ds^2 = {R^2 \over (R^2 - r^2)}dr^2 + (R^2 - r^2)d\phi^2 +
r^2d\Omega_2^2,
\ee
where $d\Omega_2^2$ is the metric of a unit 2--sphere parametrized by
$\theta_3$ and $\theta_4$. From the metric we see that the volume
element is just
\be
Rr^2drd\phi d\Omega_2.
\ee

The kinetic energy of the membrane is given by the Dirac--Born--Infeld
Lagrangian. We are mostly interested in the case when the size of the sphere $r$
is constant and close to its maximum value $R$. In this case the
membrane moves around a circle in the $X_1, X_2$ plane of radius $(R^2
- r^2)^{1/2}$ with angular
velocity $\dot \phi$. Dropping time derivatives of $r$, we have
\be
{\cal{L}}_K = -T \Omega_2 r^2 \sqrt{1 - (R^2 - r^2){\dot {\phi}}^2}.
\ee
Here, $T$ is the membrane tension which is given by
\be
T = {1 \over 4\pi^2 l_p^3}
\ee
in 11-dimensional Planck units.

Next we add the Chern--Simons coupling involving the background field.
%%\bigskip\noindent
%%{\it Coupling to the Flux}
The contribution of the four-form field strength to the action
of the brane per orbit around the $S^4$ is
\be
S_{B} = \int_{wv} C = \int_{\Sigma} F.
\ee
The first integral is over the world-volume of the brane. $F = dC$ is
the four-form flux, and $\Sigma$ is a four-manifold in the $S^4$ whose
boundary is the 3-dimensional surface swept out by the brane during one orbit.
Since the background flux is just $F = B d{\rm vol}$, where $B$ is the
constant
flux density and $d{\rm vol}$ is the volume form on $S^4$, we have
\be
S_{B} = B {\rm vol} (\Sigma).
\ee
Therefore the Chern-Simons term in the Lagrangian is
\be
{\cal L}_{B} = {S_{B} \over T} = B {\rm vol} (\Sigma)
{\dot \phi \over 2 \pi}
\ee
where $\dot \phi$ is the (constant) angular velocity of the brane.
Parametrizing the motion as above, the volume of $\Sigma$ is
\be
{\rm vol}(\Sigma) = R \Omega_2 \int_0^{2 \pi} d\phi \int_0^{r}
r'^2 dr' = {8 \pi^2 \over 3} R r^3.
\ee
So the Chern-Simons term is
\be
{\cal L}_{B} = {\dot \phi \over 2 \pi} B \Omega_4 R r^3
= \dot \phi N {r^3 \over R^3}
\ee
where we used the flux quantization condition, Eq.(3.1).

Therefore, the full bosonic Lagrangian is 
\be
{\cal L} = -m \sqrt{1 - \dot \phi^2 (R^2 - r^2)} + N {r^3 \over R^3} \dot \phi
\ee
with $ m = \Omega_2 T r^2$.
Using Eq.(3.2) and Eq.(3.15), we also see that
\be
{N \over R^3} = T\Omega_2.
\ee
From the Lagrangian we find that the angular momentum is given by
\be 
L = {m \dot \phi (R^2 - r^2) \over \sqrt{1 - \dot \phi^2 (R^2 - r^2)}} + N {r^3 \over R^3}.
\ee
The maximum size a membrane can have is $R$. Also, the velocity of its
center of mass, $\dot{\phi}R$, cannot exceed the speed of light. This
implies that the angular momentum has a maximum value given by
$N$ \footnote{There is an exception to this statement at the
pathological value $r = 0$ which we discuss at the end of this section.}:
\be
L_{max} = N.
\ee
When the membrane has maximal size $R$, the angular momentum is
the maximum value $N$.  
We see that the Kaluza--Klein graviton has a maximum angular
momentum in agreement with the stringy exclusion principle.
For the energy, we find 
\be 
E = \dot \phi L - {\cal L} = 
\sqrt{\left({Nr^2 \over R^3}\right) ^2 + {(L - Nr^3/R^3)^2 \over R^2 - r^2}}.
\ee
Varying the energy with respect to $r$ at fixed $L$, we find 
\be
{dE \over dr} = {r \over E(R^2 - r^2)^2} \left(L - N{r \over
R}\right) \left(L - 2N { r \over R} + N {r^3 \over R^3}\right).
\ee
We see that for $L<N$ there exists a stable minimum at 
\be 
r = {L \over N} R.
\ee
Therefore, the membrane grows as we increase the angular momentum.
This is in agreement with Eq.(3.4) \footnote{The cubic factor in
$dE/dr$ has a positive root at $0< r < (L/ N)R$ at which the
energy has a local maximum and another root at $r>R$.}.
When $r = R$ and $L = N$, a more careful analysis for the stability of
the solution is needed and we do so at the end of this section.
The value of the energy at the minimum is
\be
E = {L \over R},
\ee
which is the energy of a Kaluza--Klein graviton with angular momentum $L$.
From Eq.(3.23), we also find that the velocity of the center of mass
equals the speed of light, $\dot{\phi}R = 1$.

We now show that there is a stable solution at $r = R$ and $L = N$.
Setting $\tilde{r} = R - r$ and expanding the Lagrangian up to
quadratic powers in $\tilde{r}$, we obtain
\be
{\cal{L}}_K = -T \Omega_2 R^2 + T\Omega_2R \tilde{r} (2 + \dot {\phi}^2 R^2)
- T\Omega_2\tilde{r}^2 ({\dot{\phi}}^4R^4 + {5 \over 2} {\dot{\phi}}^2 R^2
+1),
\ee
and
\be
{\cal{L}}_B = \Omega_4BR^4\dot{\phi}(1 - 3{\tilde{r} \over R} +
3{\tilde{r}^2 \over R^2}) = N \dot{\phi} - R\tilde{r}{3N \over
R^3}\dot{\phi}R + \tilde{r}^2 {3N \over R^3}\dot{\phi}R  .
\ee
Using $N/R^3 = T\Omega_2$,
the total Lagrangian becomes
\be
{\cal{L}} = -T \Omega_2 R^2 + N \dot{\phi} + T\Omega_2 R \tilde{r} (2 -3
\dot{\phi}R + \dot {\phi}^2 R^2)
- T \Omega_2 \tilde{r}^2 ({\dot{\phi}}^4R^4 +
{5 \over 2}{\dot{\phi}}^2R^2 - {3} {\dot{\phi}} R +1).
\ee
There is an extremum at $\tilde{r} = 0$ provided that
\be
2 -3\dot{\phi}R + \dot {\phi}^2 R^2 = 0.
\ee
This can be achieved if the velocity $\dot{\phi}R = 1$. Thus, when the
size of the membrane is $R$ its center of mass moves with the speed of
light.
Furthermore, the extremum is stable since
\be
{d^2V(r) \over d^2r}|_{r = R} > 0.
\ee

Thus when the size of the membrane is $R$, the angular momentum has
its maximum value $N$ and the energy is given by
\be
E = T \Omega_2 R^2 = {N \over R}. 
\ee 
At the classical level, this is in exact
agreement with the energy of a Kaluza--Klein graviton
having angular momentum $N$ about the sphere. When $N>>1$, the
maximal angular momentum is large and the classical formula for the
energy agrees with the BPS bound for the energy given the angular momentum. 
Quantum corrections are suppressed by $1/N$.
The Kaluza--Klein graviton is a BPS state and its energy should not
change under the process of blowing up into a membrane.
Again, the size of the brane is determined by the angular
momentum. Since the maximum size a brane can have is $R$, there is a
maximum angular momentum as predicted by the dual conformal field theory.
The fact that the energy of the brane agrees with the BPS formula for
given angular momentum is a non--trivial test of our model.

We also note that there is a minimum of the energy at $r = 0$ as
well. Classically, it corresponds to a massless particle moving around
the equator with angular momentum $L$.\footnote{We thank Sunny Itzhaki
for discussions of this point.}
Such a solution is singular from the perspective of the gravitational
field equations since for angular momenta of order $N$, it represents
a huge energy (of order $N^{2/3}$) concentrated
at a point. Therefore it is subject to uncontrolled quantum
corrections.
In particular, there are quantum corrections proportional to powers of
the momentum times the flux density, which are large at angular momenta
of order $N$. 

We have shown in this paper that such a singular solution can be
resolved by blowing into a smooth macroscopic membrane of size
$(L/N)R$. Our classical analysis is expected to be valid for the large
membrane. The smooth membrane solution certainly has much more nearby
phase space and as a result the true quantum ground state will be
overwhelmingly supported at the membrane solution. We believe that
this is another example of how string/$M$--theory resolves
singularities of the type studied rececently in ref.\cite{21}.

\subsection{$AdS_5 \times S^5$}

The extension of our analysis to the other two maximally supersymmetric
cases is straightforward and we will be much less explicit. Consider
the case of $AdS_5 \times S^5$ first. The radius of the five sphere is
given by
\be
R = (4\pi g_s N)^{1 \over 4}l_s,
\ee
where $g_s$, $l_s$ are the string coupling constant and string
length scale and $N$ is the number of units of five--form flux on the sphere. We
take $N$ large keeping $g_s N$ fixed and large.

In type IIB string theory on $AdS_5 \times S^5$ with $N >> 1$,
the maximum angular momentum of a BPS particle on the $S^5$ is $N$
\cite{13}. From the gauge theory perspective, this can be seen from
the fact that one builds up such states by a single trace of the $N
\times N$ scalars in the $\bf 6$ of $SO(6)$.  The largest
representation of $SO(6)$ one can build in this way is the spin-$N$ 
representation, ${\rm Sym}^N {\bf 6}$.

From our perspective this is because the particle moving on the
sphere expands into a spherical D3-brane.  In this case we 
present only the exact classical analysis of the 
D3-brane wrapping an $S^3$ that moves in $S^5$.  The bosonic Lagrangian is 
\be
{\cal L} = {\cal L}_{DBI} + {\cal L}_{CS}=
-T_{D3}\Omega_3 r^3 \sqrt{1 -  (R^2 - {r^2})\dot\phi^2} + \dot \phi N {r^4
\over R^4}.
\ee
The tension of the D3-brane is
\be
T_{D3} = {1 \over (2\pi)^3 l_s^4 g_s}.
\ee  
We will use the relation
\be 
T_{D3} \Omega_3 = {N \over R^4}.
\ee

The angular momentum in terms of $\dot \phi$ is
\be 
L = {m \dot \phi (R^2 - r^2) \over \sqrt{1 - \dot \phi^2 (R^2 - r^2)}} + N {r^4 \over R^4}
\ee
where $m = T_{D3} \Omega_3 r^3 = (N/R^4) r^3$.  Again we see that the 
angular momentum is bounded by $N$ since $0 \le r \le R$ and $0 \le \dot \phi R \le 1$.
The energy is 
\be
E = \sqrt{m^2 + {(L - Nr^4/R^4)^2 \over R^2 - r^2}}.
\ee
Varying the energy with respect to $r$ at fixed $L$, we find in this case 
a stable minimum when
\be
r^2 = {L \over N} R^2.
\ee
The value of the energy at this minimum again matches the BPS bound when $L$ is
large, for $N >> 1$:
\be
E = {L \over R}.
\ee

This is strong evidence that at any appreciable 
momentum, at least at the 
(semi-) classical level, the good description
of Kaluza--Klein gravitons is in terms of branes, rather than
fundamental strings.  
From the dual CFT, we know that there is a unique BPS state 
with these quantum numbers; consistency with the exclusion principle implies that it 
is the one described by the spherical brane.

\subsection{$AdS_4\times S^7$}

In this case, we expect the graviton to expand into an M5-brane which is
an $S^5 \subset S^7$.
The radius of the sphere is given by
\be
R = (2^5 \pi^2 N)^{1 \over 6}l_p.
\ee
The tension of the 5--brane is given by
\be
T = {1 \over (2 \pi)^5 l_p^6}
\ee
and we have the relation
\be
m = T \Omega_5 r^5 = {N \over R^6} r^5.
\ee

The Lagrangian is 
\be
{\cal L} = - T\Omega_5 r^5 \sqrt{1 -  (R^2 - r^2)\dot\phi^2} + \dot \phi N
{r^6 \over R^6}.
\ee
The angular momentum in terms of $\dot \phi$ is
\be 
L = {m \dot \phi (R^2 - r^2) \over \sqrt{1 - \dot \phi^2 (R^2 - r^2)}} + N {r^6 \over R^6}.
\ee
The energy is 
\be
E = \sqrt{m^2 + {(L - Nr^6/R^6)^2 \over R^2 - r^2}}.
\ee
Varying the energy with respect to $r$ at fixed $L$, we find in this case 
a stable minimum when
\be
r^4 = {L \over N} R^4.
\ee
The value of the energy at this minimum again matches the BPS bound when $L$ is large:
\be
E = {L \over R}.
\ee

\subsection{Remarks about $AdS_3$}

We will conclude this section with some comments about
$AdS_3 \times S^3 \times M_4$. This case is distinguished
in several ways.  

First, it is not clear into what the graviton should expand.
Consider the
geometry built from the D1-D5 system with $Q_1$ D-strings and $Q_5$
five-branes.  The stringy exclusion bound on 
the angular momentum is $L \le Q_1 Q_5$.
The graviton is expected to blow up into a circular string moving on the $S^3$,
but should it be a D5-brane wrapped on the four-manifold, or a D-string?

Secondly, the energetic considerations degenerate in this case.
If we assume for argument's sake that the graviton blows up into 
either a 
D-string or wrapped fivebrane 
which is a circle on the $S^3$, we find that the energy at 
fixed $L$ has no nontrivial minimum.  Considering some 
incarnation of the ``fractional strings'' of \cite{18} does 
not help.

It may help to consider the S-dual situation of
the F1-NS5 system.  In this case, the
dynamics of fundamental strings
on the relevant sphere are described by a level-$Q_5$ $SU(2)$ WZW model.
One expects the exclusion principle to be related to the affine cutoff on
$SU(2)$ representations.  
Finally, 
a clarification of this case should match 
the result of \cite{19} that the exclusion bound occurs
at the critical value of the energy for black hole formation.
A better understanding remains for future work.

\section{Conclusions}
\setcounter{equation}{0}

Physics in non-commutative spaces is characterized by a simple
signature -- the increase of size on systems  with increasing
momentum. In this paper we have seen that the motion of massless
quanta on the $S$ factor of $AdS_n \times S^m$ has exactly this
behavior. The massless particle blows up into a spherical brane of
dimensionality $m-2$ whose radius increases with increasing
momentum. Eventually the radius of the blown up brane becomes
equal to the radius of the sphere that contains it. It can no
longer grow and the spectrum is terminated. This is the origin of
the stringy exclusion principle. Thus we see one more piece of
evidence for non-commutativity of space in quantum gravity.

\section{Acknowledgments}
 
We would like to thank Allan Adams, Micha Berkooz, Raphael Bousso, 
Sumit Das, Antal Jevicki, Jason Prince Kumar,
Albion Lawrence, Alec Matusis, and Joe Polchinski for conversations.  We are
especially grateful to Joe Polchinski for explaining to us the
flat-space energetics.  We thank Hirosi Ooguri and Sunny Itzhaki for 
comments about the first version.
We would also like to thank Steve Shenker for 
help with the title.  The work of 
JM is supported in part by the Department of Defense NDSEG 
fellowship program.  The work of LS and NT is 
supported in part by NSF grant 9870115.

%%%%%%%%%%%%%%%%%%%%%%%%%%%%%%%%%%%%%%%%%%%%%%%%%%%%%%%%%%%%%%%%%%%%%%%%%%%%
%%%%%

%%%%%%%%%%%%%%%%%%%%%%%%%%%%%%%%%%%%%%%%%%%%%%%%%%%%%%%%%%%%%%%%%%%%%%%%%%%%
%%%%%
%%%%%%%%%%%%%%%%%%%%%%%%%%%%%%%%%%%%%%%%%%%%%%%%%%%%%%%%%%%%%%%%%%%%%%%%%%%%
%%%%%
\end{document}